\documentclass[conference]{IEEEtran}
\IEEEoverridecommandlockouts
\usepackage{cite}
\usepackage{amsmath,amssymb,amsfonts}
\usepackage{algorithmic}
\usepackage{graphicx}
\usepackage{textcomp}
\usepackage{xcolor}
\usepackage{listings}
\usepackage{lineno}
\usepackage[colorinlistoftodos]{todonotes}
\usepackage{balance}

\def\BibTeX{{\rm B\kern-.05em{\sc i\kern-.025em b}\kern-.08em
    T\kern-.1667em\lower.7ex\hbox{E}\kern-.125emX}}
\begin{document}

\title{MPro: Combining Static and Symbolic Analysis for Scalable Testing of Smart Contract\\
}

\author{\IEEEauthorblockN{William Zhang, Vijay Ganesh}
\IEEEauthorblockA{
\textit{University of Waterloo}\\
Waterloo, Canada \\
\{william.zhang,vijay.ganesh\}@waterloo.ca}
\and
\IEEEauthorblockN{Sebastian Banescu, Leonardo Passos, Steven Stewart}
\IEEEauthorblockA{\textit{Quantstamp Inc., Blockchain Development Labs} \\
San Francisco, California, USA \\
\{sebi,leo,steven\}@quantstamp.com}
}

\maketitle

\begin{abstract}


Smart contracts are executable programs that enable building a programmable trust mechanism between multiple entities without the need for a trusted third-party. At the time of this writing, over 10 million smart contracts were deployed on the Ethereum networks, and this number continues to grow at a rapid pace. Smart contracts are often written in a Turing-complete programming language called Solidity, which is not easy to audit for subtle errors. Further, since smart contracts are immutable, errors have led to attacks resulting in losses of cryptocurrency worth 100s of millions of USD and reputational damage. Unfortunately, manual security analyses do not scale with the size and number of smart contracts. Automated and scalable mechanisms are essential if smart contracts are to gain mainstream acceptance. Researchers have developed several security scanners in the past couple of years. However, many of these analyzers either do not scale well or, if they do, produce many false positives. This issue is exacerbated when bugs are triggered only after a series of interactions with the functions of the contract-under-test. A \emph{depth-n vulnerability}, refers to a vulnerability that requires invoking a specific sequence of $n$ functions to trigger. Depth-n vulnerabilities are time-consuming to detect by existing automated analyzers because of the combinatorial explosion of sequences of functions that could be executed on smart contracts.


In this paper, we present a technique to analyze depth-n vulnerabilities in an efficient and scalable way by combining symbolic execution and data dependency analysis. A significant advantage of combining symbolic with static analysis is that it scales much better than symbolic alone and does not have the problem of false positive that static analysis tools typically have. We have implemented our technique in a tool called MPro, a scalable and automated smart contract analyzer based on the existing symbolic analysis tool Mythril-Classic and the static analysis tool Slither. We analyzed 100 randomly chosen smart contracts on MPro, and our evaluation shows that MPro is about $n$-times faster than Mythril-Classic for detecting depth-$n$ vulnerabilities while preserving all the detection capabilities of Mythril-Classic.

\end{abstract}


\section{Introduction}
The Ethereum Blockchain, with a capitalization of 14.8 billion USD \cite{price} and more than 59 million existing accounts \cite{number}, is one of the leading platforms for smart contracts. 
Unfortunately, smart contracts remain vulnerable to various security attacks. 
Several attacks have been observed in the past few years like the DAO attack in 2016, which resulted in the theft of 60 million USD worth of Ether \cite{DAO}.
The Parity Wallet attack in 2017 led to losses of over 30 million USD due to improper access control of a sensitive function, which could be called through a publicly accessible function \cite{parityhack}.
Another highly publicized attack was a \emph{batch overflow} attack in 2018, which led to the creation of very large amounts of several different crypto-currency tokens \cite{batchoverflow}. 
This attack caused reputation damage for the affected crypto-currency projects as well as for Ethreum as seen by the drop of 8\% in the ETH value on the day this attack was announced.

Symbolic execution \cite{symex} is a promising technique adopted widely in smart contract analyzers, which tries to explore all possible program execution paths and uses constraint solvers to calculate a concrete input (test case) for each feasible path. However, as the possible program states and paths grow exponentially with the number of branches in the program, the scalability of this technique is often inhibited due to the path explosion \cite{PathExploision} issue. 
Nevertheless, the often limited size of smart contracts is a blessing, which allows symbolic execution to be highly applicable in this domain, as opposed to the domain of classical software systems, where real-world software nowadays has millions of lines of code.

One of the most popular security scanners for smart contracts, \emph{Mythril-Classic} \cite{Mythril}, uses symbolic execution at its core.
It is one of the most comprehensive and powerful tools among all the existing symbolic analyzers, which models the \emph{Ethereum Virtual Machine} (EVM) \cite{EVM} to a high degree of accuracy and supports a wide range of security checks. 


Another issue that current security scanners (including Mythril-Classic) face is the combinatorial explosion of different possible function call sequences of arbitrary lengths in smart contract interactions. 
Depth-n vulnerabilities in smart contracts were first presented in \cite{MAIAN} as \textit{trace vulnerabilities}, which refers to the vulnerability that cannot be triggered in a single function invocation. 
To detect depth-n vulnerabilities, Mythril-Classic currently uses brute force (with a basic path pruning strategy) to execute all possible sequences of function calls up to the maximum call depth, which often leads to combinatorial explosion. 
Our experiment of 100 smart contracts presented in Section~\ref{sec:evaluaiton}, shows that 34\% of the test cases contain depth-n vulnerabilities. The overall runtimes of Mythril-Classic at depth 2 and depth 3 are 3.43x and 14.04x longer than the overall execution time at depth 1. 
This result highlights the importance of optimizing the symbolic execution engine of Mythril-Classic at deeper depths.

In this paper, we present an optimization technique for detecting depth-n vulnerabilities based on data dependency analysis and we integrate it with Mythril-Classic. 
By leveraging data dependency analysis, unnecessary sequences of function invocations can be pruned during the analysis. 
Function call sequences that are more likely to lead to depth-n vulnerabilities are prioritized (and therefore depth-n vulnerabilities can be detected in an early stage), based on the read/write dependencies of state variables in every externally accessible function of the smart contract. 

\textbf{This paper makes the following contributions:}
\begin{enumerate}
\item An optimization based on data-dependency analysis, to the problem of finding depth-n vulnerabilities via symbolic execution.
\item An open-source implementation of this optimization approach based on Mythril-Classic, called MPro\footnote{https://github.com/QuanZhang-William/M-Pro}.
\item An evaluation of this approach using MPro and a dataset of 100 real-world smart contracts.
\end{enumerate}

The rest of this paper is organized in the following way.
Section~\ref{sec:background} presents the background knowledge necessary to understand the rest of this paper.
Section~\ref{sec:design} describes the design of our optimization approach and the implementation of MPro.
Section~\ref{sec:evaluaiton} presents the setup and results of our evaluation of MPro.
Section~\ref{sec:limitations} describes the limitations of our implementation.
Section~\ref{sec:rel-work} presents related work and compares it to MPro.
The conclusions and future work are presented in section~\ref{sec:conclusion}.

\section{Background}
\label{sec:background}
\subsection{Ethereum and Smart Contracts}
A smart contract, in essence, is a program stored on the Ethereum blockchain and is identified by a unique contract address.
The Ethereum blockchain is maintained and updated by many nodes connected in a peer-to-peer manner to the Ethereum network.
Each node on the Ethereum network runs the EVM.
Users can interact with the Ethereum network in order to: (i) create new contracts; (ii) invoke functions of a contract; (iii) transfer the cryptocurrency Ether (ETH) to other contracts or users. 
A function invocation to a smart contract is recorded inside a \emph{transaction}.
The Ethereum blockchain can also be described as a transaction-based state machine, where the sequence of transactions on the blockchain determines the state of each smart contract and the balance of each user ~\cite{MythrilPaper}. 
Each smart contract contains its private storage, a predefined executable bytecode and may control some Ether. 
When a user invokes a function of a smart contract, the EVM \cite{EVM}, executes the bytecode corresponding to that function, which is often compiled from Turing-complete programming languages like Solidity or Vyper.
The following section presents different types of vulnerabilities that arise due to the complexity of these programming languages and the underlying programming paradigm.

\subsection{Smart Contract Security}
Different from traditional software programs, a smart contract is publicly accessible, transparent and immutable. 
This leads to situations where bugs, including security issues, are visible and exploitable by attackers. 
Attackers are highly incentivized to exploit such vulnerabilities for their own benefit as smart contracts often hold some amount of crypto-currency. 
In this section, we specifically describe two types of vulnerabilities that can be detected by MPro (and Mythril-Classic). 
A complete list of detection capabilities of Mythril-Classic and MPro can be found in \cite{Mythril}.

\paragraph{Reentrancy}
This is a well-known vulnerability residing in smart contracts as it resulted in the DAO attack in 2016\cite{DAO}. 
Figure~\ref{fig:reentracy_code_example} shows a simple smart contract that is vulnerable to reentrancy.

\lstset{
    language=C,
    captionpos=b,
    tabsize=3,
    frame=none,
    keywordstyle=\color{blue},
    commentstyle=\color{gray},
    stringstyle=\color{red},
    numbers=left,
    numberstyle=\footnotesize,
    numbersep=5pt,
    breaklines=true,
    showstringspaces=false,
    basicstyle=\ttfamily\scriptsize,
    emph={label},
    moredelim=[is][\underbar]{__}{__},
    keepspaces=true,
    xleftmargin=.2in,
    xrightmargin=.1in
}

\begin{figure}
\begin{lstlisting}
contract Caller {
    mapping (address => uint) public balance;
    
    function withdraw(uint amount) public{
        if (balance[msg.sender] >= amount){
            require(msg.sender.call.value(amount)());
            balance[msg.sender]-=amount;
        }
    }
    
    function withdrawFixed(uint amount) public{
        if (balance[msg.sender] >= amount){
            balance[msg.sender]-=amount;
            require(msg.sender.call.value(amount)());
        }
    }
}
\end{lstlisting}
\caption{Ethereum smart contract vulnerable to reentrancy attacks}
\label{fig:reentracy_code_example}
\end{figure}

In Solidity, a smart contract can invoke a function from another smart contract using the \texttt{call} instruction and such calls are synchronous. 
In other words, the caller smart contract always waits for the callee contract to finish before executing the next line of code.  
The purpose of the \textit{withdraw} function (line 4 in Figure~\ref{fig:reentracy_code_example}) is to send out a specific amount of Ether to an external account (withdrawer) iff the withdrawer has enough credit stored in the current contract, indicated by the global variable \textit{balance}. 
The balance of the withdrawer is updated after sending Ether out (line 7 in Figure~\ref{fig:reentracy_code_example}). 

A potential reentrancy attack can happen in line 6. 
The external contract being called on line 6 can simply call back to the \textit{withdraw} function of the caller contract again. 
Note that the state variable \textit{balance} is not yet updated at this point because the caller contract is still waiting for the external call to finish. 
Therefore, in the case of a reentrancy attack, the \emph{true} branch on line 5 will be executed again and again; hence all the Ether stored in the caller contract can be drained out after a couple of iterations.

One of the simplest ways to prevent such an attack is to call an external function only after all the persistent storage updates are done \cite{BestPractice}. Thus, to fix this vulnerability, the \textit{balance} state variable needs to be updated before executing the external call (as indicated in the \textit{withdrawerFixed} function starting on line 11). Since invoking a single function is sufficient to trigger this specific vulnerability, we classify this specific reentrancy vulnerability as a \emph{depth-1 vulnerability}. 
However, it is important to note that not all reentrancy vulnerabilities are  \emph{depth-1 vulnerabilities}.

\paragraph{Unrestricted Suicide Vulnerability} 
A smart contract remains alive on the Ethereum blockchain until it gets killed by the contract owner or other authorized agents by invoking the \texttt{selfdestruct} instruction in Solidity. However, arbitrarily killing a smart contract can result in Ether lock or unexpected Ether flow, and therefore the \textit{selfdestruct} instruction should always be protected from being invoked by any arbitrary caller of the smart contract.

Figure~\ref{fig:suicide} shows a simple smart contract that can be destroyed by any arbitrary caller. In this example, the \textit{kill} function on line 13 is protected by the \textit{onlyOwner} modifier (defined on line 4), which only allows the smart contract owner (stored in the \textit{owner} field) to access the function. However, any arbitrary caller can become the owner by simply calling the \textit{setOwner} function(which is not protected).

To exploit the vulnerability, an attacker merely needs to invoke the \textit{setOwner} function (on line 9 of Figure~\ref{fig:suicide}) to become the owner of the contract and then kill the contract by calling the \textit{kill} function. 
This vulnerability was discovered in the Parity wallet in 2017 and it caused losses of 30 million USD worth of Ether \cite{parityhack, parity}. 
In this paper, we call vulnerabilities that require a specific function invocation sequences to trigger as \emph{depth-n vulnerability}, where n is the minimal number of function calls required to trigger the vulnerability. 
Thus, the unrestricted suicide issue in this example is a \emph{depth-2 vulnerability}. 
It is worth mentioning that this vulnerability is only detected by MPro (and Mythril-Classic) at transaction depth 2, but not at transaction depth 1.

\begin{figure}
\begin{lstlisting}
contract Suicide {
    address public owner;

    modifier onlyOwner{
      if(msg.sender != owner) revert();
      _;
    }

    function setOwner() public{
        owner = msg.sender;
    }

    function kill(address addr) public onlyOwner{
        selfdestruct(msg.sender);
    }
}
\end{lstlisting}
\caption{A smart contract with unrestricted suicide vulnerability}
\label{fig:suicide}
\end{figure}

\section{Design and Implementation}
\label{sec:design}
This section presents the dependency analysis technique, the heuristic symbolic engine and the design of MPro in depth.

\subsection{Data Dependency Analysis}
The \emph{key observation} of this paper is that at least one state variable dependency is necessary between a sequence of function calls in order to cause a depth-n vulnerability. 
As one transaction (function call) completes, changes to local variables (stored on the stack and in memory) are discarded and only state variables changes will be reflected in the following transactions, because they are located in persistent storage. 
In other words, a smart contract does not contain depth-n vulnerabilities if no state variable dependency exists. 
Furthermore, we investigated the eleven pre-defined security analysis modules of Mythril-Classic and found that all the depth-n vulnerabilities that can be detected by Mythril-Classic are caused by \textit{Read After Write (RAW)} function dependencies of a smart contract.

To implement data dependency analysis in MPro we use \emph{Slither}, because it provides convenient APIs to instantly analyze variable usages in each function of the smart contract \cite{Slither}.
\emph{Slither} is a highly scalable static analysis tool which analyzes a smart contract source code at the intermediate representation \textit{SlithIR} level.
In this paper, we use a slightly modified version of Slither to analyze the state variable usage in every function of the target smart contract. 
Based on the state variable usage returned by Slither, we determine the prioritized transaction sequences in the following manner:
\begin{enumerate}
\item For every externally accessible function \textit{f} in the given contract, find all the state variables read in \textit{f}.
\item For every state variable \textit{v} that is read in function \textit{f}, find all the functions in the given contract that write to the state variable.
\end{enumerate}

The \textit{RAW dependency} is therefore determined for function \textit{f}. The function dependency information is stored in a dictionary, where the key is every externally accessible function and the value is a set of corresponding functions that the key function depends on. Note that a function can have \textit{RAW dependency} to itself if a state variable is both read and written in the function.

\subsection{Mythril-Classic Workflow}
Both Mythril-Classic and MPro consist of two main components: (1) the symbolic execution engine and (2) the eleven predefined security analysis modules. 
One or more smart contract security properties are encapsulated in each of those analysis modules. 
At a high-level, the tool first invokes its symbolic execution engine, executes the target contract symbolically and produces: an execution tree representing all possible program states, execution paths and constraints. 
During the symbolic execution, the executed program states and constraints are passed to the security analysis modules when necessary. 
The generated path constraints and the \textit{negation} of security property constraints of a security module are together solved by the Z3 SMT solver. 
A violation of the safety property (i.e., a vulnerability) is reported if the overall constraint is \textit{satisfiable}.
The details of security analysis modules are not specifically explained in this paper as MPro does not enhance the security analysis modules of Mythril-Classic.

\subsection{Original Mythril-Classic Symbolic Engine}
\begin{figure}
\includegraphics[height=2.2in, width=3.5in]{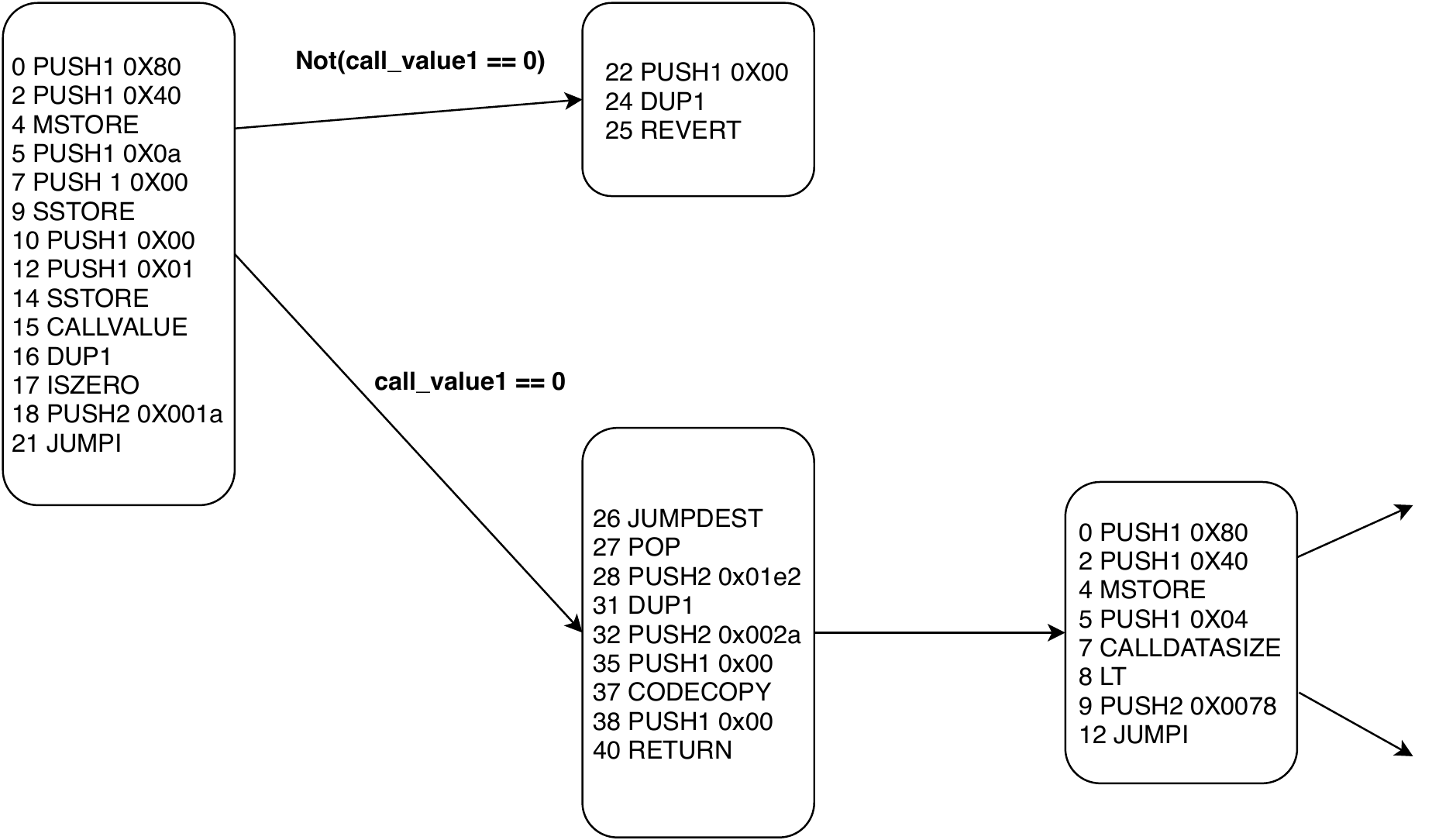}
\caption{Partial execution tree generated by Mythril-Classic}
\label{fig:workflow}
\end{figure}

Since the visualization of full execution trees would not fit on one page (such that it is still readable), Figure~\ref{fig:workflow} shows an example of an execution sub-tree generated by Mythril-Classic. 
Each node consists of consecutive EVM opcodes \cite{EVM}, representing a basic block of the program. 
A node may end with a termination instruction (i.e.~\texttt{RETURN}, \texttt{STOP}, \texttt{REVERT}, \texttt{ASSERT\_FAIL}, \texttt{SUICIDE}) or a jump instruction (i.e.~\texttt{JUMPI}, \texttt{JUMP}). 
Each instruction in a node, in essence, represents a state change of the smart contract. 
Every node is associated with a pointer to the next basic code block (except for termination instructions), and therefore, a path is constructed. 
Every path carries a path constraint showing the reachability of the state. 
All the previous executed program states are cached. 
When executing a new opcode, the previous program state is retrieved, and a new program state is determined based on the previous state and the current instruction. 
The modified new states are then pushed back to the cache and will be served as the initial states for the next instruction execution.

To detect a depth-n vulnerability, Mythril-Classic leverages brute force (with a basic path pruning strategy) to model all possible transaction sequences. 
To be more specific, the symbolic engine takes every end state of the previous transaction as an initial state to the next transaction and continues the next symbolic execution iteration. 
Mythril-Classic uses two types of path pruning techniques and the follow-up transaction execution is pruned if either: 
\begin{enumerate}
\item The end state of the previous transaction is terminated with either the \texttt{REVERT} opcode (which means all the program state changes are reverted) or the \texttt{SUICIDE} opcode (which means the smart contract is no longer alive on blockchain).
\item There is no write to the persistent storage of the smart contract in the previous transaction (as the state persistent storage remains the same).
\end{enumerate}
However, these techniques are not sufficient to mitigate the combinatorial explosion issue of Mythril-Classic when faced with smart contracts with many functions that can be called in several different orders to form a myriad of possible transaction sequences.

\subsection{MPro Heuristic Symbolic Engine}
The heuristic symbolic engine of MPro is built on top of the original symbolic engine in Mythril-Classic, providing additional features to prune unnecessary execution paths (function sequences) while preserving the original path pruning techniques. 
Our heuristic symbolic engine does not modify the execution of code blocks inside a function. 
Instead, it guides the path branching to the next function in the corresponding dependency list (computed by Slither) when the previous function terminates, and skips unnecessary path generation.

The MPro heuristic symbolic engine works in 4 steps:
\begin{enumerate}
\item Invoke the Mythril-Slither interface, retrieve and store \textit{the RAW function dependencies} in a look-up table.

\item Use the original Mythril-Classic symbolic execution engine, symbolically executes all the functions to call depth 1.

\item For all terminating states at call depth 1, mark the state as the starting state for the next transaction only if the function in the previous transaction is in the set of keys of the look-up table, which means the function in the previous transaction has a \textit{RAW dependency} to at least another function.

\item When determining the function call in the next transaction, look-up the dependency list, guiding the original symbolic engine to execute the desired branch that executes the function call in the next transaction depending on the previous transaction and skips other branches.
\end{enumerate}

\begin{figure}
\includegraphics[height=2.2in, width=3.5in]{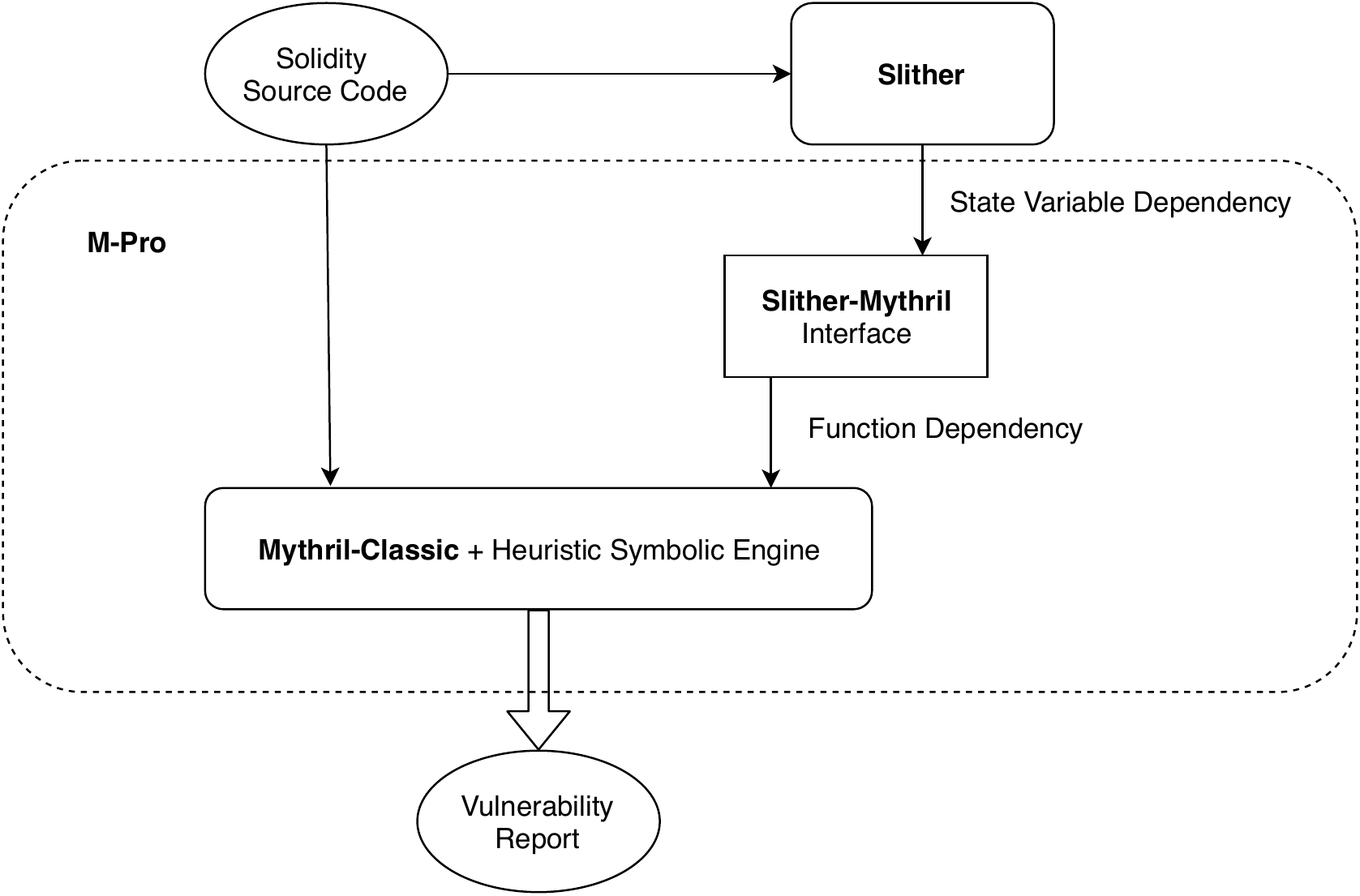}
\caption{MPro Architecture}
\label{fig:MPro_Workflow.pdf}
\end{figure}

\subsection{Design of MPro}
Figure~\ref{fig:MPro_Workflow.pdf} illustrates the workflow of MPro. It takes a smart contract source code (in Solidity) as input to Slither and Mythril-Classic. Slither first statically analyzes the state variable usages in the given smart contract and then passes the usage information to the Mythril-Slither interface, where the \textit{RAW function dependency} is analyzed.

The original solidity source code and the dependency information are then taken as inputs to MPro's heuristic symbolic execution engine. The engine executes the function combinations in a determined sequence based on the given dependency information. The heuristic symbolic engine in MPro is constantly guided along the path to execute the function combinations that have \textit{RAW dependency}. Other paths are treated as unnecessary function sequences and are therefore pruned from the search space.

\section{EVALUATION}
\label{sec:evaluaiton}

\subsection{Experimental Setup} 
We evaluated the performance and the vulnerability reports of \emph{MPro} and \emph{Mythril-Classic v0.20.0} (the latest version of Mythril-Classic at the time writing the paper) on 4 Microsoft Azure  \emph{Standard D2s v3} instances, each with 2 virtual CPUs and 8 GB memory. We used \emph{solc v0.4.25} and \emph{solc v0.5.4} for the smart contracts that require version 0.4 and 0.5 of the compiler respectively. The constraint solver used is \emph{Z3 v4.8.4.}.

We evaluate MPro and Mythril-Classic on 2 datasets: (1) the SWC test suite \cite{SWC} with 102 simple smart contract that contains vulnerabilities (and some fixes) and 100 real-world smart contracts with an average of 258.76 LOC.
The set of real-world contracts was constructed in the following way: 32 contracts were submitted (for a fee) to the Quantstamp Smart Contract Scanning protocol\footnote{https://quantstamp.com/}. The remaining 68 smart contracts from our dataset were taken from the Ethereum Mainnet and Ropsten as the latest deployed contracts at the time when we collected our dataset (i.e.~end of January 2019), with the condition that their source code is available and unique in our dataset.

The goals of our measurement are: (1) to validate that the vulnerability reports of MPro and Mythril-Classic are identical and (2) that MPro is more efficient than Mythril-Classic.

\subsection{Security}
As the security analysis modules of MPro remain the same as Mythril-Classic, we did not specifically check false positives or false negatives of the issues detected by either MPro or Mythril-Classic. This is not in the scope of this paper. Instead, we examined the vulnerability reports of MPro for both datasets and validated that MPro produces identical vulnerability reports as Mythril-Classic at depths 1, 2 and 3, given no timeout after which the tools would be stopped. 

We note that if a timeout (e.g.~30 minutes) was used for MPro and Mythril-Classic, then MPro was able to detect more vulnerabilities for some contracts, than Mythril-Classic at depths 2 and 3. More details are shown in the next section. In terms of the security evaluation for this paper, we can say that MPro is at least as good as Mythril-Classic. We leave the improvement of the security capabilities of Mythril-Classic for future work.

\subsection{Performance}
We measure the performances of MPro and Mythril-Classic with the 100 randomly chosen real-world smart contracts. 
The performance evaluation of the SWC test suite is skipped as the test cases in SWC test suite use the minimum amount of code to showcase a specific kind of vulnerability, they are not real-world contracts. 

\begin{table}
\caption{Performance Comparison (depth 2)}
\begin{center}
\begin{tabular}{|c|c|c|c|c|c|}
\hline
\textbf{Timeout}&\multicolumn{2}{c|}{\textbf{\textit{MPro}}}& \multicolumn{2}{c|}{\textbf{\textit{Mythril-Classic}}}& \textbf{\textit{Speedup}} \\
\cline{2-5}
& Avg.~time(s) & T-out & Avg.~time(s) & T-out & \\
\hline
10-min & 116 & 8 & 167 & 17 & 1.61x  \\
30-min & 182 & 3 & 309 & 5 & 2.04x  \\
none & 176 & 0 & 301 & 2 & 1.92x  \\
\hline
\end{tabular}
\label{table1}
\end{center}
\end{table}

\begin{table}
\caption{Performance Comparison (depth 3)}
\begin{center}
\begin{tabular}{|c|c|c|c|c|c|}
\hline
\textbf{Timeout}&\multicolumn{2}{c|}{\textbf{\textit{MPro}}}& \multicolumn{2}{c|}{\textbf{\textit{Mythril-Classic}}}& \textbf{\textit{Speedup}} \\
\cline{2-5}
& Avg.~time(s) & T-out & Avg.~time(s) & T-out & \\
\hline
120-min & 856 & 12 & 1262 & 20 & 2.89x\\
\hline
\end{tabular}
\label{table2}
\end{center}
\end{table}

\paragraph{Overall Speedup and Timeouts}
We measured the performance of the two tools with and without a timeout at depths 1, 2 and 3.
At depth-1, both tools have approximately the same performance, minus a small delta needed by MPro to run the data dependency analysis. 
However, it would be a trivial optimization to exclude this analysis if the user is interested only in depth-1.

At depth-2, we used timeouts of 10- and 30-minutes and noticed that 9 and 3 more programs respectively timed-out in Mythril-Classic than in MPro, where 8 and 5 programs respectively timed-out.
Table~\ref{table1} summarizes the performance comparison between MPro and Mythril-Classic at depth 2. 
On average, MPro almost reaches a 2x speedup comparing to Mythril-Classic with no timeout.
Given a timeout of 10 minutes, the speedup is lower because the execution time of the timed-out instances is capped at 10 minutes. 
However, given this 10 minutes timeout, more than double the number of programs timeout in Mythril-Classic, in comparison to MPro.
In addition, two instances are killed by the operating system when running Mythril-Classic with no timeout due to memory restriction. 
The same two instances finish successfully in MPro, because of our path pruning strategy, which leads to less memory usage.


At depth-3, we used a higher timeout of 120-minutes to be fair, and we noticed that eight more programs timed-out in Mythril-Classic than MPro, where 12 out of 100 programs timed-out.
Table~\ref{table2} summarizes the performance comparisons at depth 3, where we can see that at depth-3, the speedup of MPro w.r.t.~Mythril-Classic is almost 3x.
Similarly, to the observation for depth-2, given a timeout, the speedup is lower; however, the number of timed-out instances for Mythril-Classic is almost double that of MPro.
We stop the evaluation at depth 3 as no depth-3 vulnerability is detected by either tool for our dataset.
Note that we do not report the average time at depth-3 for the analysis with no timeout because we ran the 20 programs that timed out on Mythril-Classic for over two days before manually stopping the analysis process. Hence we cannot accurately compute the speedup when there is no timeout, but we not that it would likely be even higher than 2.89x that was recorded for a timeout of 120 minutes.

\paragraph{Speedup Correlated with Smart Contract Complexity}
We further investigate the test cases that the performance of MPro, particularly dominates Mythril-Classic at depth 2 and 3. We put the test cases into two categories: the test cases that finish between 0 to 10 minutes in Mythril-Classic, the test cases that take more than 10 minutes to finish.
Table~\ref{table3} shows the speedup of MPro of the two categories. 
As the complexity of the smart contract increases (indicated by the overall runtime of Mythril-Classic), the MPro speedup reaches 4.45x at depth 2 and 6.51x at depth 3, with test cases that take more than 10 minutes to finish in Mythril-Classic. 
This trend is expected to continue, i.e.~the more complex a smart contract is, the bigger the overall execution tree; hence more of the search space can be pruned by the data dependency analysis in MPro.

\begin{figure}
\includegraphics[height=2.2in, width=3.5in]{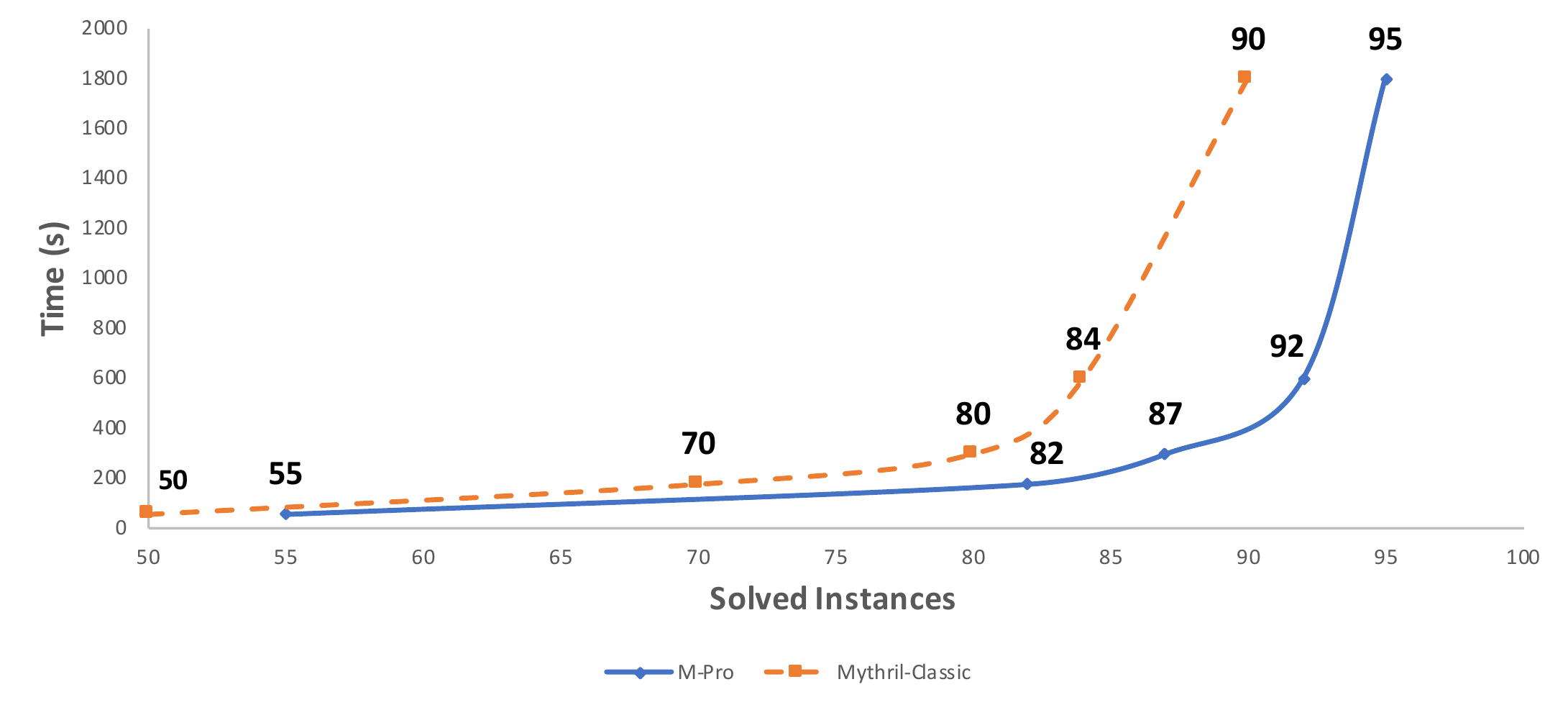}
\caption{Comparison of Solved Instances at depth 2}
\label{fig:catcus_D2.pdf}
\end{figure}

\begin{figure}
\includegraphics[height=2.2in, width=3.5in]{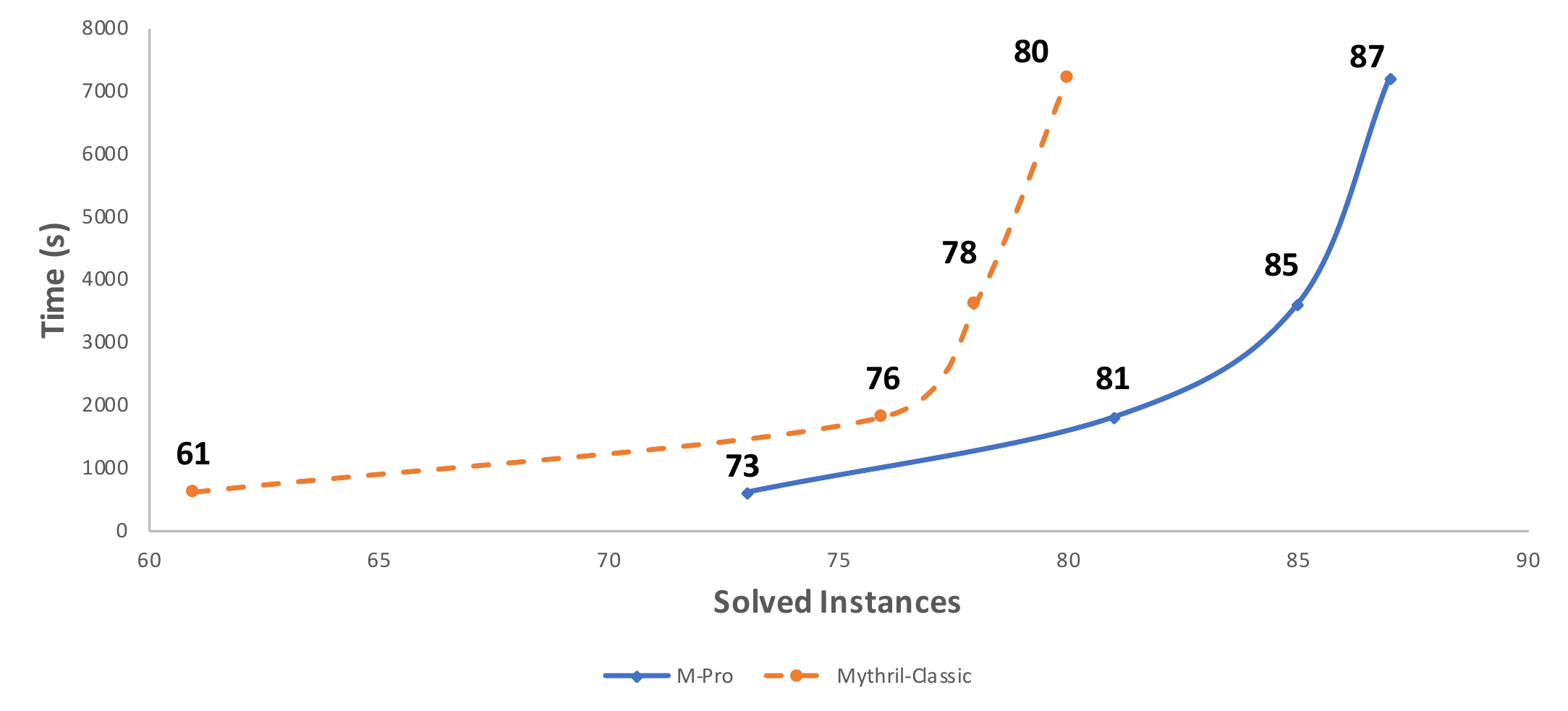}
\caption{Comparison of Solved Instances at depth 3}
\label{fig:catcus_D3.pdf}
\end{figure}

\paragraph{Number of Solved Instances}
In addition to the performance speedup, we also measure the advantage of MPro in terms of the number of instances solved in a certain amount of time.
Specifically, we sampled the number of test cases from the 100 real-world test suite that finished in 60, 180, 300, 600, and 1800 seconds in both tools. 
Figure~\ref{fig:catcus_D2.pdf} demonstrates such comparison at depth 2 where the x-axis is the number of instances solved from the 100 randomly chosen smart contract test suit, and the y-axis is the time period. 
As shown in the graph, MPro has an absolute advantage over Mythril-Classic, where MPro is able to solve 10\%, 17.14\%, 8.75\%, 9.52\%, and 5.56\% more instances than Mythril-Classic at the corresponding sampling point. 
Similarly, we record the number of solved instances at depth 3 for both tools at 600, 1800, 3600, and 7200 seconds threshold respectively, and MPro is able to solve 19.67\%, 6.57\%, 8.97\% and 8.75\% more instances than Mythril-Classic as shown in Figure~\ref{fig:catcus_D3.pdf}.

\begin{table}
\caption{Performance Comparison Categorized by Complexity}
\begin{center}
\begin{tabular}{|c|c|c|c|c|}
\hline
\textbf{Depth}&\textbf{Exec.~Time}&\textbf{\textit{MPro (s)}}& \textbf{\textit{Mythril-Classic (s)}}& \textbf{\textit{Speedup}} \\

\hline
2 & t $\leq$ 10 mins & 52 & 79 & 1.40x \\
2 & t $\geq$ 10 mins & 783 & 1383 & 4.45x  \\
\hline
3 & t $\leq$ 10 mins & 94 & 125 & 1.78x \\
3 & t $\geq$ 10 mins & 1206 & 3087 & 6.51x  \\
\hline
\end{tabular}
\label{table3}
\end{center}
\end{table}

\section{Limitations}
\label{sec:limitations}
\emph{Limitations.} 
MPro has several limitations, which we describe in this section.
First, the data dependency analysis of Slither is not control-flow sensitive. Specifically, Slither simply reports all the state variables that show up in a function. 
However, some state variables may not be reachable. 
This may lead to an imprecise function dependency analysis in MPro and hence non-dependent function combinations during symbolic execution.
Fixing this limitation would require adopting a more precise static analysis technique, which we intend to pursue in future work.

Second, MPro does not optimize the intra-procedural symbolic execution of the functions in smart contracts, but only heuristically prunes unnecessary follow-up functions. 
According to the performance evaluation of MPro, we believe that there is still room for improvement since there are still some real-world smart contracts that time out in 30 or 120 minutes in MPro.
We intend to address this issue in future work.

Finally, since MPro uses Slither for the data dependency analysis, it requires the source code of the smart contract as input.
This limitation will be eliminated in future work by employing a data dependency analysis technique directly on the smart contract bytecode.

\section{Related Work}
\label{sec:rel-work}
In this section, we present related smart contract analyzers. 
Since in this paper, we mainly focus on performance optimization and not on detecting more types of vulnerabilities or reducing false positives, we limited our evaluation to a comparison with Mythril-Classic.
A direct comparison of the performance of the following tools and MPro is not valid, because they detect different types of vulnerabilities.

\emph{Maian} \cite{MAIAN} (no longer being maintained) is a symbolic analyzer for smart contracts which is able to model multiple transactions of a smart contract. 
The Maian team first defined a formal and systematic characterization of \textit{trace vulnerabilities}. 
By leveraging concolic analysis, Maian initializes partial program states with concrete values, which makes the project scalable to analyze real-world smart contract to some extent. 
To further increase scalability, Maian utilizes basic static analysis to filter out unnecessary symbolic engine usage. 
Maian can detect only three vulnerable patterns: Greedy, Prodigal, and Suicidal. 

\emph{Securify} is a fully automated and static Ethereum smart contract analyzer. 
Securify defines two kinds of patterns for a security property: \textit{compliance patterns}, which implies the satisfaction of the property, and \textit{violation patterns}, which implies its negation \cite{Securify}. 
Thanks to the patterns, the tool is able to produce some definite violations of given security properties, which reduces the manual effort required to verify the issue detected. 
Securify provides 7 built-in security properties and corresponding compliance/violation patterns like "no write after call", "restricted write" and "restricted transfer". 

\emph{Zeus}\cite{ZEUS} is a sound and scalable framework for automated formal verification of smart contracts based on abstract interpretation and model checking. 
Zeus classifies the issues in smart contracts into two categories: correctness issues and fairness issues. 
Zeus analyzes the input smart contract at a low-level intermediate representation and allows users to customize the correctness/fairness criteria in a given template. 

\section{Conclusion and Future Work}
\label{sec:conclusion}
We have presented MPro, a fully automated and scalable security scanning tool for Ethereum smart contracts and we offer it as free open source software. 
The tool is developed based on Mythril-Classic and Slither, leveraging both static and symbolic analysis to prune unnecessary execution paths. 
We have evaluated the efficiency of MPro and compared it to Mythil-Classic using a dataset of 100 real-world smart contracts that we compiled ourselves and which we also offer to the general public.
From our experimental observations, MPro is on average $n$ times more efficient than Mythril-Classic while preserving all the detection capabilities of Mythril-Classic when detecting depth-$n$ vulnerabilities.

\emph{Future Work.} Going further, we propose two major improvements of MPro: (\textit{i}) to develop a Slither-driven targeted symbolic mechanism \cite{targetedSE} based on the MPro heuristic symbolic engine, and (\textit{ii}) to parallelize the symbolic execution and security analysis processes of Mythril-Classic, to further improve the performance of MPro.
Since the calculation of a concrete input to validate the detected vulnerability is the main limitation of static tools like Slither, we plan to solve this problem by implementing a more sophisticated mechanism which further combines static analysis with a symbolic engine. 
In the Slither-driven targeted symbolic execution mechanism, MPro is aimed at generating test cases to prove the presence of exploitable vulnerabilities by targeting the warnings flagged by Slither. 
The general idea is to guide the symbolic execution engine in MPro along all the branches that are more likely to hit a target vulnerable state flagged by Slither. 
The potential vulnerability is proved if a concrete input can be returned by the symbolic engine and the constraint solver. 
Otherwise, the detected issue can be classified as a false positive.

Since the current version of MPro only guides the branches to the next dependent transaction, this improvement can be treated as a technique to prune search spaces on top of MPro further. 
Ideally, all other paths that do not lead to an error can be avoided. 
As Slither is efficient and a single path generation is cheap, this mechanism is expected to be more efficient.

In addition, the analyses of both Mythril-Classic and MPro are sequential. While invoking multiple instances of a symbolic engine to parallelize constructing the one overall execution tree often leads to redundancies, it is feasible to split the overall execution tree into multiple subtrees by the initial function calls. 
Each tree holds only one initial function call and its dependents. 
The parallel execution of such independent subtrees is relatively easy.

\end{document}